**The First Human-Based In Vitro Flow Loop and Quantification for Fetal Aortic Hemodynamics**

Kamlesh Kumar[1]; Biao Si[1]; Pritom Saha[1]; Shuping Ge[2]; Zhenglun Alan Wei[1,3,4]

[1]Department of Biomedical Engineering, Worcester Polytechnic Institute, Massachusetts, USA

[2]Department of Pediatric and Adult Congenital Cardiology, Geisinger Heart and Vascular Institute, Geisinger Medical Center, Danville, USA

[3]Department of Mechanical Engineering, Worcester Polytechnic Institute, Worcester, MA, USA

[4]Department of Data Science, Worcester Polytechnic Institute, Worcester, MA, USA

Please send the correspondence of the current work to

Name: Zhenglun Alan Wei, PhD

Address: 60 Prescott Street

Phone: 978-934-3754

E-mail: ZWei1@wpi.edu

Word Limit = 6500,

Word Count, Total = 6775.5

- Word: 4317
- Figures:
  - single column figure 200 × 6 = 1200
  - double column figure 400 × 3 = 1200
- Tables, single column 7 lines × 6.5 + 13 = 58.5

Highlight the manuscript text, excluding abstract, author list, acknowledgments, and references, making note of the word count at the bottom of the screen. Add to that the word-count equivalents for figures, tables, and equations as follows:

- **Figures:** An average single-column figure will displace 200 words; an average double column figure will displace 400 words
- **Tables:** 6.5 words per line, plus 13 words for single-column tables. 13 words per line, plus 26 words for double-column tables.
- **Equations:** 7 words per line for single-column equations. 13 words per line for double-column equations.

If the total number of words (text + figures + tables + equations) is 6125 or less, the length is acceptable.

**ABSTRACT (WORD LIMIT: 250; WORD COUNT: 242)**

Coarctation of the aorta (CoA) is a common congenital defect that remains difficult to diagnose prenatally due to subtle and evolving anatomical features. In the fetus, the ductus arteriosus creates a dual-inflow configuration that generates complex three-dimensional flow patterns not captured by standard imaging. Improved characterization of fetal hemodynamics may enhance diagnostic accuracy beyond anatomy-based assessment.

This study presents the first human-based in vitro flow loop of the fetal aorta, constructed from anatomies reconstructed using medical imaging data. Models representing normal and coarctation conditions were fabricated and integrated into a physiological flow loop. Velocity fields were measured using planar and stereoscopic particle image velocimetry (PIV) to resolve near-wall and three-dimensional flow structures, enabling quantitative assessment of velocity gradients and wall shear stress (WSS) under normal and coarctation configurations.

The in vitro flow loop closely reproduced target fetal flow segmentation, with segmental flow-rate errors generally below 6%. High-resolution planar and stereoscopic PIV revealed dual jets from the ascending aorta and the ductus arteriosus and predominantly planar flow in the normal aorta, but strong jet acceleration, separation, and reattachment in the coarcted geometry. Coarctation produced markedly elevated and spatially heterogeneous WSS, and 2-component PIV underestimated WSS by up to ~29% compared with 3-component measurements, especially in high-shear regions.

These findings show that accurate three-component velocity measurements are critical for reliable WSS estimation and suggest that detailed hemodynamic metrics, such as WSS, may serve as potential biomarkers to enhance fetal CoA diagnosis beyond anatomy alone.

## A. INTRODUCTION

Congenital heart disease affects approximately 1% of live births.[1, 2] Among these conditions, coarctation of the aorta (CoA) is one of the most prevalent defects and is characterized by a localized narrowing of the aortic isthmus (AoI), typically just proximal to the descending aorta (DAo).[3-8] In the fetal circulation, the aortic arch exhibits distinct hemodynamic features due to the presence of the ductus arteriosus (DA), a transient vascular connection that closes after birth. The DA shunts blood from the pulmonary artery (PA) to the DAo,[9-11] fundamentally altering the distribution of flow and pressure within the aortic arch. This configuration creates a dual-inflow system in which flow from the DA merges with flow from the ascending aorta (AAo) near the AoI, producing asymmetric momentum exchange and complex three-dimensional (3D) flow structures.

Despite its clinical importance, prenatal diagnosis of CoA remains challenging. Fetal anatomical markers are often subtle and can evolve throughout gestation, resulting in high rates of both false-positive and false-negative diagnoses.[12] Current clinical assessment relies primarily on echocardiographic anatomical measurements, yet diagnostic accuracy for isolated CoA frequently remains below 50%.[13] These limitations highlight the need for complementary approaches that incorporate hemodynamic information.

From a fluid mechanics perspective, the fetal aortic arch can be viewed as a curved, branching conduit with a localized constriction. This geometric feature accelerates flow, suppresses boundary-layer development, and increases velocity gradients, thereby elevating vascular stresses. The resulting flow is characterized by momentum redistribution and jet formation in the AoI, which in turn generate spatially heterogeneous velocity and shear fields.[14, 15] Capturing these features requires spatially resolved measurements that characterize both bulk flow structures and near-wall gradients.[16-18]

Such detailed characterization is clinically relevant. Patients with native and repaired CoA exhibit persistent abnormal helical and eccentric flow patterns throughout the thoracic aorta, even after successful intervention, indicating long-term alterations in aortic hemodynamics. These flow abnormalities are associated with altered wall shear stress (WSS) distributions and adverse vascular remodeling.[19, 20] In addition, near-wall hemodynamics, particularly spatial variations in WSS, play a central role in endothelial function and vascular adaptation. In postnatal CoA, disturbed shear environments have been linked to hypertension, vessel wall remodeling, and long-term cardiovascular complications.[21-28] Therefore, accurate quantification of both near-wall flow gradients and global flow structures is essential for understanding fetal CoA hemodynamics and improving diagnostic evaluation.

Current clinical tools for assessing fetal aortic hemodynamics include echocardiography and four-dimensional flow magnetic resonance imaging (4D-flow MRI).[29-35]

- Ultrasound remains the primary modality due to its real-time capability and widespread availability; however, it cannot resolve detailed spatial velocity fields and provides limited information on near-wall flow behavior and WSS.[36]
- 4D-flow MRI enables volumetric visualization of flow and offers global hemodynamic assessment,[37] but its spatial resolution near the vessel wall is insufficient for accurate estimation of velocity gradients and WSS.[37]

Image-based computational fluid dynamics (CFD) offers a powerful alternative, enabling detailed predictions of velocity fields, vortex dynamics, and WSS in complex fetal vascular geometries.[2, 38] However, CFD results depend strongly on modeling assumptions, geometric reconstruction accuracy, and prescribed boundary conditions. Consequently, high-resolution experimental data are required for validation before CFD can be reliably translated into clinical practice,[38] yet such validation remains scarce for fetal aortic hemodynamics.

In vitro physiological flow loops offer a well-established platform for studying cardiovascular fluid mechanics, validating CFD models, and evaluating device performance.[39-41] While extensive studies have investigated aortic hemodynamics in pediatric and adult populations,[17, 35, 42, 43] only one prior study has combined an in vitro flow loop with CFD to examine the fetal aortic arch under both normal and pathological conditions.[10] However,

- that work primarily relied on idealized geometries and qualitative flow visualization;
- although CFD revealed complex 3D flow phenomena, quantitative assessment of near-wall velocity gradients and wall shear stress (WSS) was not performed in either the CFD or the in vitro experiments.

To address these gaps, there is a clear need for a physiologically realistic, human-based in vitro platform capable of quantifying fetal aortic hemodynamics. In this study, we develop a first-of-its-kind image-based fetal aortic flow loop derived from medical imaging data. Particle image velocimetry (PIV) is employed to quantify flow fields in both normal and CoA conditions. Planar PIV (2D2C) and stereoscopic PIV (2D3C) are used to provide, for the first time, quantitative characterization of 3D flow structures in the fetal aortic arch.

## B. MATERIALS AND METHODS

An experimental platform was developed to investigate fetal aortic hemodynamics using a human-based in vitro flow loop under steady-flow conditions. The system integrates a subject-specific fetal aortic phantom (Section B.1), a physiologically representative flow loop (Section B.2), and a high-resolution particle image velocimetry (PIV) system capable of both planar (2D2C) and stereoscopic (2D3C) measurements (Section B.3). A schematic of the overall setup is shown in **Fig. 1**.

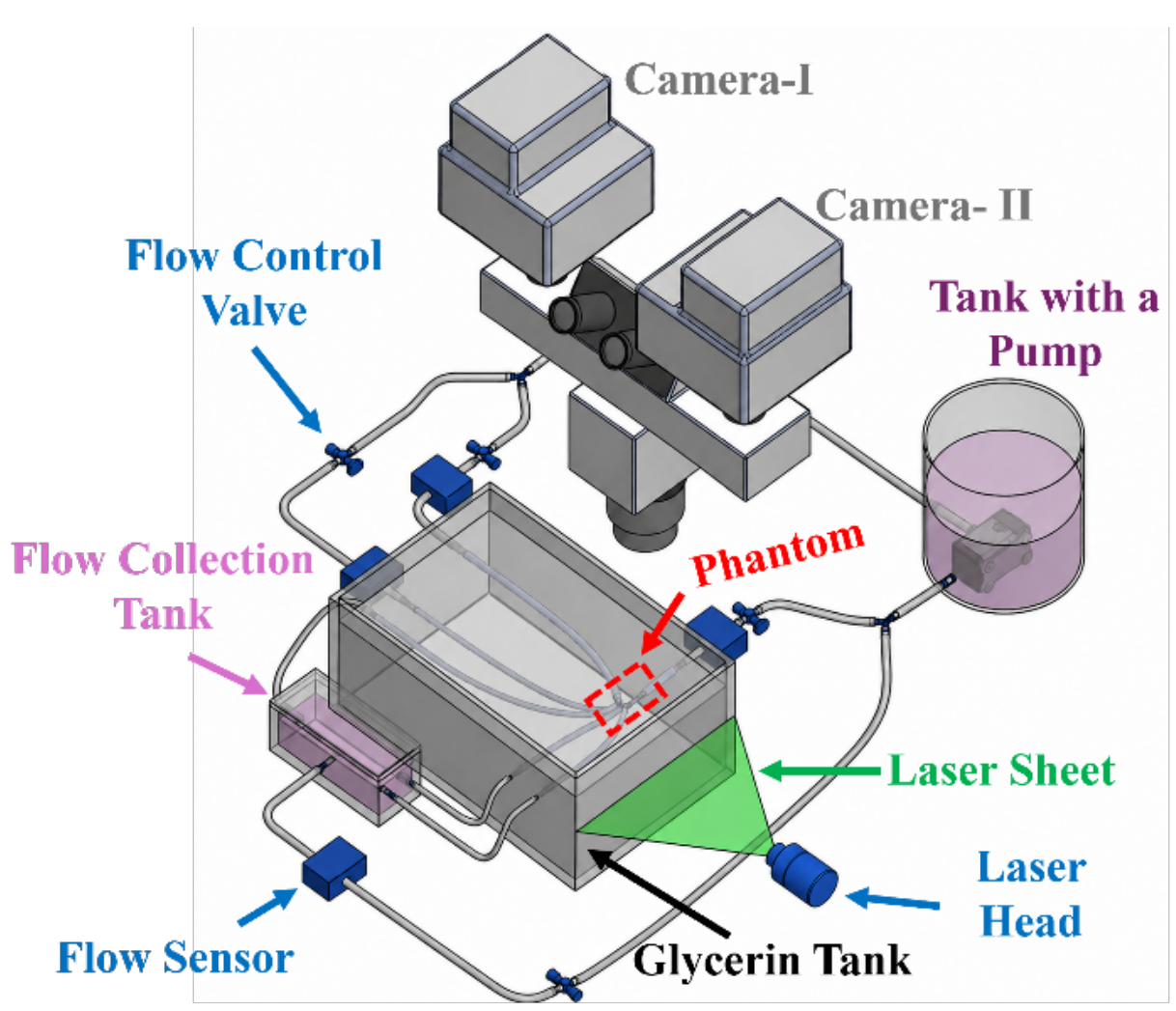

Figure 1 Schematic of the human-based in vitro experimental setup for the fetal aorta.

### B.1 Human-based Fetal Aortic Phantom

#### B.1.1 Anatomical Reconstruction and Flow Segmentation

A normal human fetal aorta at 35.5 weeks of gestation was analyzed in this study. The anatomical model was reconstructed in our previous work using a validated imaging pipeline.[2] Imaging data were acquired using a clinical ultrasound system (GE Voluson E10) with 3D fetal echocardiography and Doppler ultrasound, yielding 30 slices with a spatial resolution of 0.5 mm.[2] Flow information was obtained from Doppler measurements by tracking the spectral envelope over the cardiac cycle using a smoothed contouring approach.[2] Time-averaged velocities were extracted and used to define steady inflow conditions for the experiments.

### B.1.2 Phantom Fabrication

The reconstructed geometry (**Fig. 2**) was fabricated using stereolithography (SLA) 3D printing (Formlabs Form 4) with a layer thickness of 100 µm. The model was oriented at 45° relative to the build platform to balance surface quality and structural stability while minimizing support artifacts in the region of interest (RoI).[44] A wall thickness of 1 mm was selected to ensure mechanical integrity. The printer's dimensional tolerance is ±0.15% of nominal geometry.[45] Post-processing included isopropyl alcohol cleaning and ultraviolet curing. To enable accurate optical measurements, both internal and external surfaces were polished to improve transparency and reduce refractive distortion (**Fig. 2c–d**). This process was time-intensive but essential for achieving high-quality PIV imaging. Material removal was limited to surface smoothing and did not significantly alter lumen geometry.

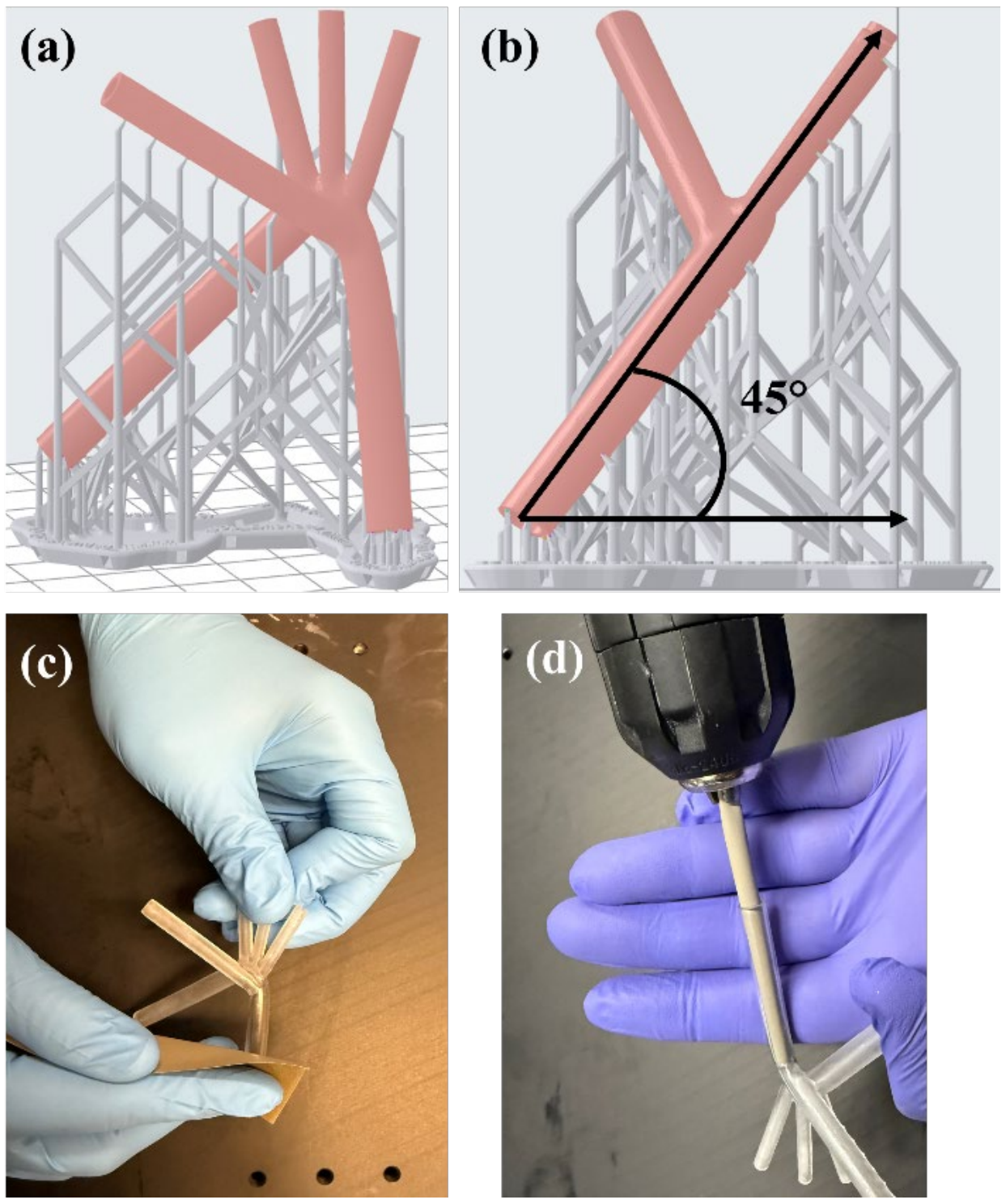

Figure 2 CAD model of the human-based fetal aortic phantom and surface finishing process. (a) Front view and (b) side view illustrating the 45° build orientation during stereolithography; (c) external sanding and (d) internal sanding used to improve optical transparency and reduce surface roughness for PIV measurements

A coarctation model was fabricated using the same process, incorporating a localized narrowing at the AoI while preserving global arch geometry and ductal connectivity (**Fig. 3**). Identical fabrication conditions ensured that observed flow differences are attributable solely to geometric variation. The refractive index of the phantom material was measured using Snell's law by recording incident and refracted laser angles, ensuring optical matching with the working fluid.

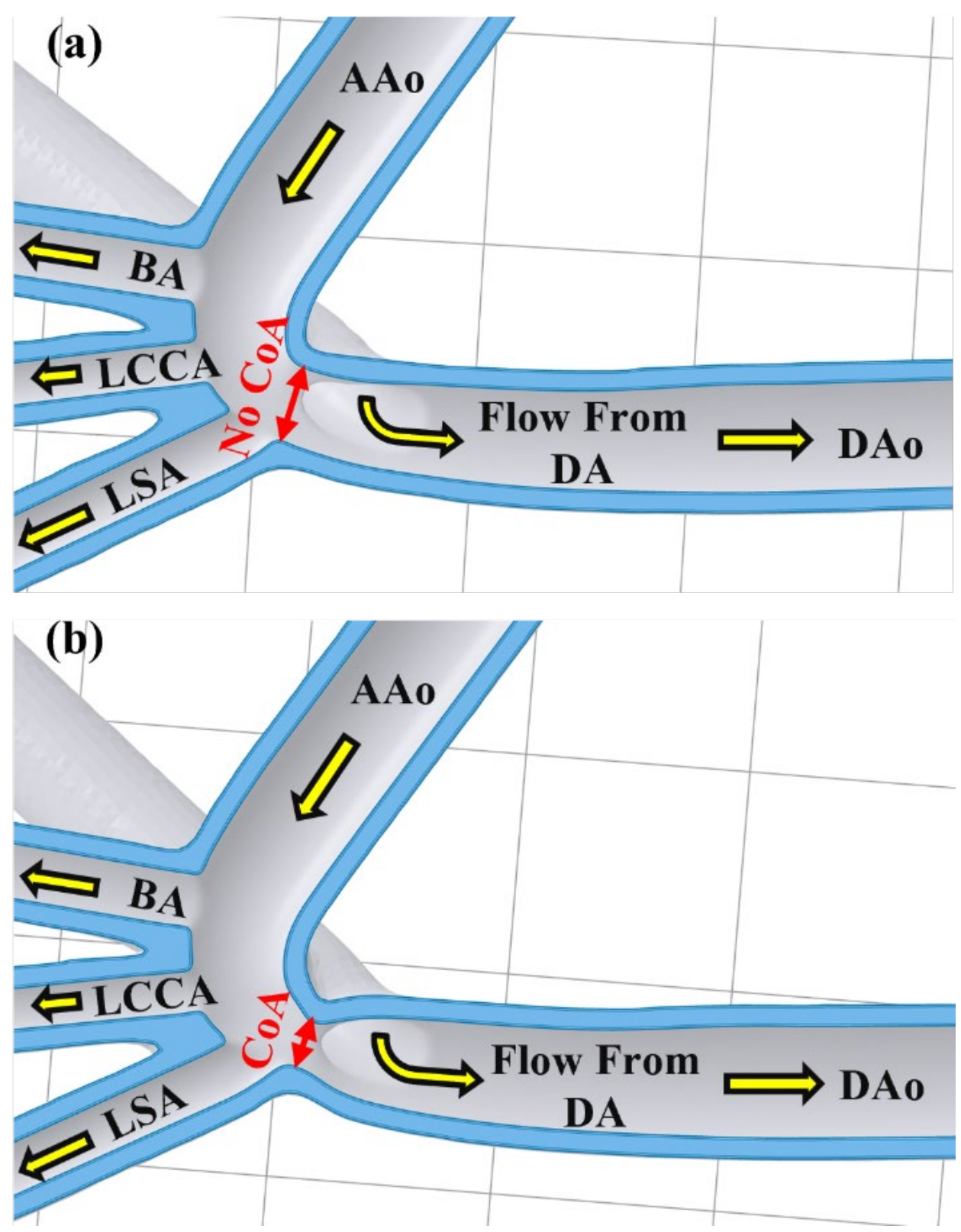

Figure 3 Section view of the human-based fetal aortic CAD models. (a) Normal aortic arch geometry. (b) The coarctation model is derived from normal anatomy and incorporates a localized narrowing of the aortic isthmus just proximal to the ductus arteriosus. AAo: ascending aorta; BA: brachiocephalic artery; CoA: Coarctation of the Aorta; DA: ductus arteriosus; DAo: descending aorta; LCCA: left common carotid artery; LSA: left subclavian artery.

### B.2 Physiological Flow Loop

A closed-loop experimental system was constructed to generate controlled steady flow within the fetal aortic phantom (**Fig. 1**). The phantom was submerged in a tank filled with a blood-mimicking fluid, and flow was supplied through both the AAo and DA to reproduce the dual-inflow configuration of the fetal circulation. Four outlets were incorporated (as shown in **Fig. 3**): the descending aorta (DAo) and three supra-aortic branches: brachiocephalic artery (BA), left common carotid artery (LCCA), and left subclavian artery (LSA). Outflows from the supra-aortic branches were routed to a collection reservoir before recirculation, whereas the DAo outlet was connected directly to the main reservoir; due to sensor limitations, only the combined outlet flow rate was measured.

The RoI was located downstream of the AoI within the proximal DAo (as shown in **Fig. 3**), where the laser sheet was aligned for velocity measurements. Steady-flow conditions were used to isolate spatial flow features and enable accurate quantification of near-wall gradients, and all experiments were repeated under identical conditions to verify reproducibility of the measured velocity fields.

A blood-mimicking fluid (BMF) was prepared to match the rheological properties of fetal blood. Based on gestational age, the target dynamic viscosity was approximately 0.0035 Pa·s.[2] Viscosity measurements were performed using a rotational rheometer (MCR 302, Anton Paar) with a cone-plate geometry (CP50 1/TG) at 23°C, which corresponds to the experimental conditions.

### B.3 Particle Image Velocimetry (PIV)

Planar PIV was used to measure two-component velocity fields. The flow was seeded with fluorescent particles and illuminated using a dual-pulse Nd:YLF laser (527 nm, 90 W at 3 kHz), forming a laser sheet approximately 0.5 mm thick. Images were acquired using a high-resolution CMOS camera (Phantom VEO E340) coupled with a Zeiss V20 microscope. Image pairs were recorded at 800 frames per second with a resolution of 2560 × 1600 pixels. Calibration yielded a spatial resolution of 0.005 mm/pixel. The inter-frame time ($\Delta t$ = 30μs) was selected to achieve particle displacements of approximately 6 pixels. Velocity vectors were computed using a multi-pass cross-correlation algorithm with a final interrogation window of 24 × 24 pixels and 75% overlap, yielding a spatial resolution of 0.12 mm and a vector spacing of 0.029 mm. This resolution is sufficient for near-wall gradient estimation.

Stereoscopic PIV was performed to obtain all three velocity components. Two synchronized cameras were positioned at angles of approximately 30–35° relative to the laser sheet.[46] A multi-plane calibration procedure was used to reconstruct the out-of-plane velocity component.[46] Measurements were conducted under identical conditions as planar PIV, enabling direct comparison and assessment of 3D flow effects, particularly those arising from curvature-induced secondary motion.

Fluorescent particles with diameters of 10–20 μm were used as tracers. The Stokes number was calculated and found to be much less than 1, indicating negligible particle inertia and accurate tracking of the flow (Appendix A).[47-49]

Wall shear stress (WSS) was computed using an in-house algorithm based on near-wall velocity gradients derived from PIV data. A locally linear velocity profile was assumed near the wall, and gradients were estimated to use velocity samples at a defined wall-normal offset. An offset of approximately half the interrogation window size provided stable WSS estimates and was adopted for all analyses.[50, 51] The algorithm for calculating WSS is detailed in Appendix B, and the uncertainty analysis and comparison with a validated computational model are presented in Appendix C.

## C. RESULTS AND DISCUSSION

### C.1 Fabrication, Optical Fidelity, and Working Fluid Characterization

The 3D-printed phantom successfully preserved the intended subject-specific geometry and exhibited uniform structural integrity after post-processing (**Fig. 4**). **Figure 4(a)** shows the model with support structures in place, while **Fig. 4(b)** presents the phantom after support removal, with the RoI and the lower wall (used for WSS calculation) highlighted by a blue dotted line.

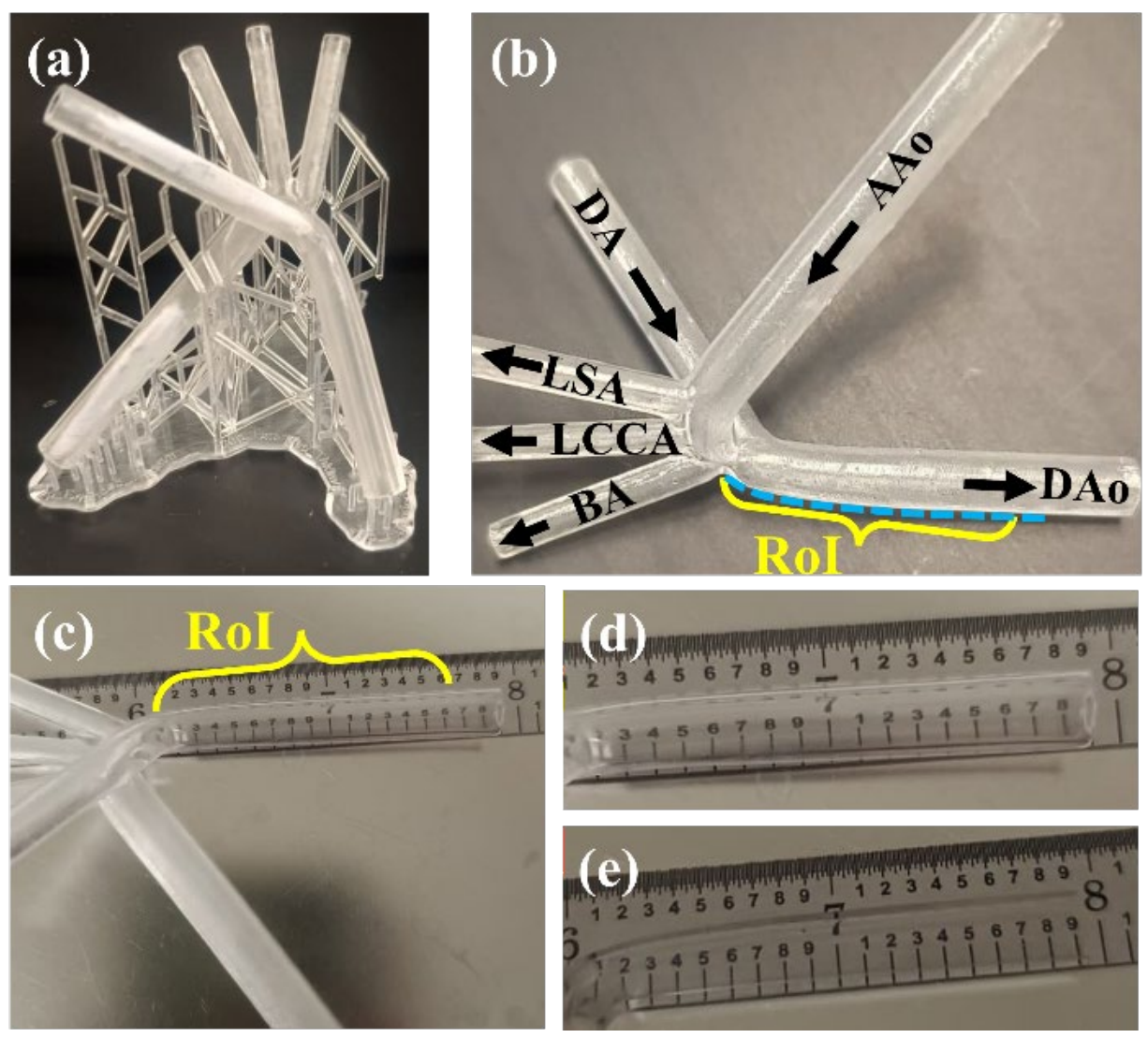

Figure 4 3D-printed human-based fetal aortic phantom and optical assessment. (a) Phantom with support structures. (b) Phantom after support removal, showing the region of interest and lower wall (blue dashed line) used for the calculation of wall shear stress. (c) Final processed model highlighting the region of interest. (d) Magnified view demonstrating improved surface transparency. (e) Phantom immersed in a glycerin bath, illustrating reduced optical distortion and effective refractive index matching. AAo: ascending

aorta; BA: brachiocephalic artery; DA: ductus arteriosus; DAo: descending aorta; LCCA: left common carotid artery; LSA: left subclavian artery; RoI: region of interest.

Following surface finishing, the phantom achieved sufficient optical transparency for PIV measurements (**Fig. 4c-e**). **Figure 4(c)** illustrates the final processed phantom and identifies the RoI used for PIV imaging, whereas **Fig. 4(d)** provides a magnified view of this region. A ruler placed behind the model demonstrates minimal light scattering through the wall, providing a qualitative assessment of transparency. When the phantom was immersed in a glycerin bath with a refractive index of approximately 1.47, only minor refraction was observed, indicating effective refractive index matching with the working fluid. This is consistent with the measured refractive index of the printed material (= 1.52). Together, these results confirm that the optical configuration and working fluid provide suitable conditions for PIV, enabling reliable velocity and near-wall shear measurements in the curved, contracting fetal aortic geometry.

The blood-mimicking working fluid, consisting of a glycerin–water mixture (36% glycerin and 64% water by volume), exhibited a measured dynamic viscosity of $\mu = 0.0035$ Pa·s at 23°C, matching the target fetal blood viscosity specified in the Methods section. Overall, the combination of geometric fidelity, optical clarity, and well-characterized fluid properties establishes a robust experimental framework for resolving velocity fields and near-wall shear in the present study.

### C.2 Physiological Flow Loop and Tuning

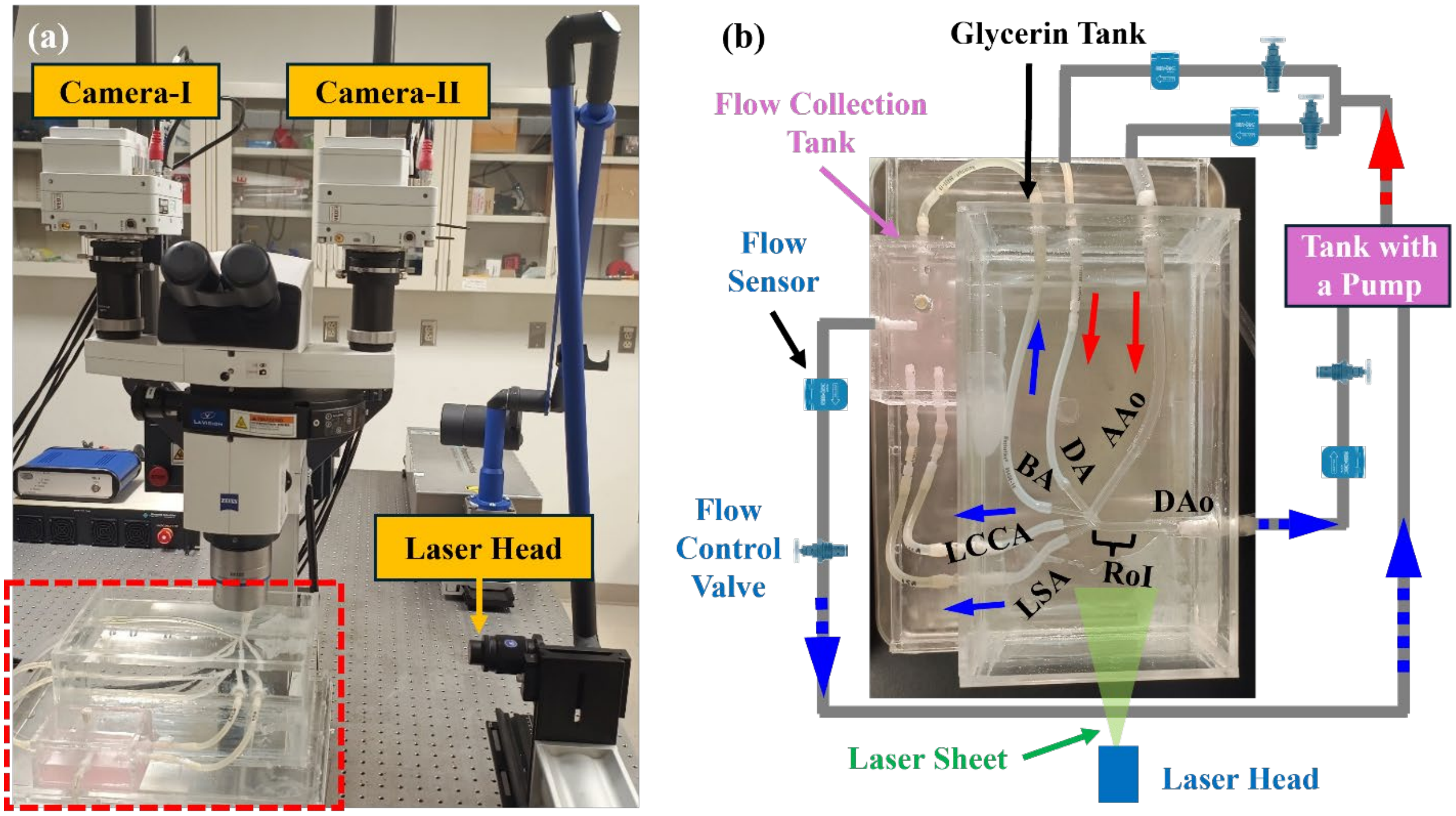

Figure 5 Experimental particle image velocity (PIV) setup and in vitro flow loop. (a) Overview of the PIV system with the fetal aortic phantom loop highlighted by a red dashed box. (b) Close-up of the closed-loop configuration showing the phantom submerged in the refractive-index–matched tank and annotated inlet and outlet connections. AAo: ascending aorta; BA: brachiocephalic artery; DA: ductus arteriosus; DAo: descending aorta; LCCA: left common carotid artery; LSA: left subclavian artery; RoI: region of interest.

The experimental PIV system and integrated flow loop are shown in **Fig. 5**. The overall PIV setup is presented in **Fig. 5(a)**, with the fetal aortic phantom flow loop highlighted by a red dotted box, while **Fig. 5(b)** details the closed-loop configuration, including the phantom submerged in the glycerin tank and the associated connections and components.

Volumetric flow distribution within the loop closely reproduced the target fetal flow segmentation (**Table 1**). The maximum deviation between prescribed and measured flow rates was 5.44% in the arch branch region (BA + LCCA + LSA), whereas all other vessels showed deviations below 3.5%. The smallest deviation (0.39%) occurred at the AoI, indicating excellent agreement along the primary flow pathway. Flow at the AoI was evaluated using mass conservation: an upstream estimate ($Q_{AoI_upstream}$) was obtained by subtracting branch outflows from AAo flow, and a downstream estimate ($Q_{AoI_downstream}$) was computed from the difference between DAo flow and DA inflow, i.e., $Q_{AoI_upstream}$=AAo−(BA+LCCA+LSA) and $Q_{AoI_downstream}$=DAo−DA. The close agreement between $Q_{AoI_upstream}$ and $Q_{AoI_downstream}$ confirms the internal consistency of the flow distribution within the loop. Additionally, the Reynolds number at the descending aorta (≈ 850–900) indicates laminar flow conditions.

Overall, the good agreement between target and measured flow rates demonstrates that the experimental setup accurately reproduces the intended fetal flow distribution, providing a reliable basis for subsequent velocity field and WSS analyses.

Table 1 Comparison of clinical (target) and in vitro flow rates in fetal aortic segments. AAo: ascending aorta; BA: brachiocephalic artery; DA: ductus arteriosus; DAo: descending aorta; LCCA: left common carotid artery; LSA: left subclavian artery; RoI: region of interest.

| Vessel Name | Clinical, mL/min | In Vitro, mL/min | Error (%) |
|---|---|---|---|
| **AAo** | 597 | 605 | 1.3 |
| **DA** | 199 | 206 | 3.5 |
| **DAo** | 603 | 596 | 1.2 |
| **BA+LCCA+LSA** | 193 | 204 | 5.7 |
| $\mathbf{Q_{AoI_upstream}}$ | 404 | 401 | 0.7 |
| $\mathbf{Q_{AoI_downstream}}$ | 404 | 390 | 3.5 |

## C.3 Three-Dimensional Flow Structure and Velocity Profiles

### C.3.1 Flow Field in the Normal Fetal Aorta and its 3D Effect

Velocity fields in the RoI were obtained using both 2D2C and 2D3C PIV (**Fig. 6**). The 2D2C configuration resolves the in-plane velocity components, capturing the dominant streamwise transport, whereas 2D3C additionally reconstructs the out-of-plane component. The resulting contours reveal two distinct jet-like structures: a primary jet originating from the AAo and a secondary jet arising from the DA.

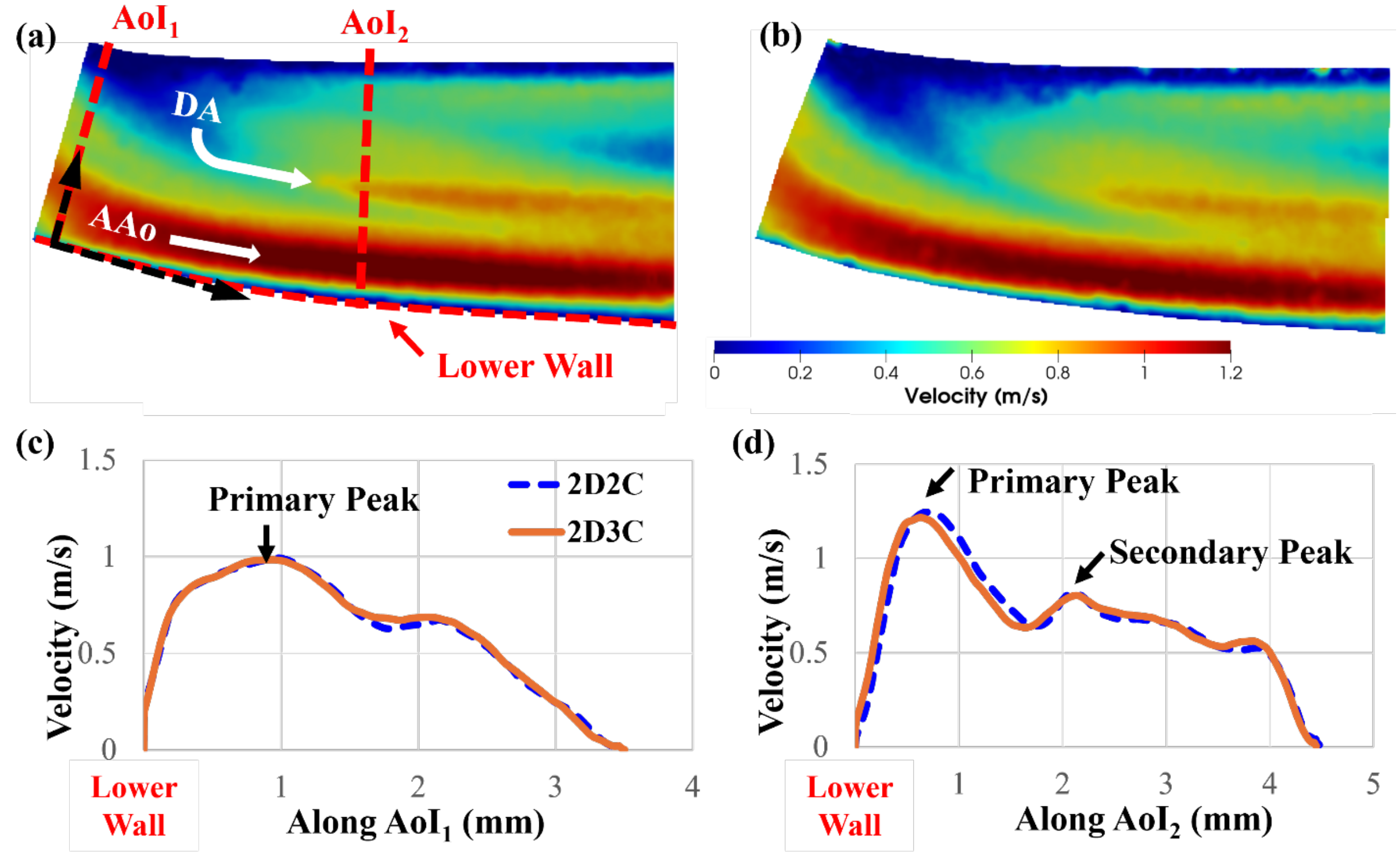

Figure 6 Velocity field in the normal fetal aortic arch within the region of interest obtained from (a) 2D2C and (b) 2D3C PIV, showing primary AAo and secondary DA jets. Velocity profiles at cross-section lines, (c) $AoI_1$ and (d) $AoI_2$, highlight the emergence of a secondary peak downstream of the DA junction and demonstrate that the flow is predominantly planar with weak 3D effects. AAo: ascending aorta; AoI: aortic isthmus; DA: ductus arteriosus.

Two cross-section lines, $AoI_1$ and $AoI_2$, are defined upstream and downstream of the DA junction, respectively, to quantify the influence of the secondary jet. The lower wall is identified as the minimum wall-normal position within the measurement plane, and a local tangent–normal coordinate system is defined at the lowest point along the $AoI_1$, coinciding with the lower wall [**Fig. 6(a)**] and serving as the reference for WSS evaluation. These definitions are applied consistently across all cases to enable direct comparison between normal and coarcted geometries.

For the normal aorta, peak velocities are concentrated near the lower wall. At the cross-section line $AoI_1$, a single dominant peak is observed, associated with the AAo inflow. At $AoI_2$, the interaction between the DA jet and the primary stream produces a secondary peak, yielding a non-monotonic profile with multiple local extrema. The 2D3C contours [**Fig. 6(b)**] exhibit a similar in-plane structure and comparable velocity magnitudes to 2D2C [**Fig. 6(a)**], indicating that planar measurements capture the primary momentum distribution. However, 2D3C resolves a finite, though relatively small, out-of-plane velocity component that persists across RoI.

Velocity profiles at the cross-section lines ($AoI_1$ and $AoI_2$ in **Fig. 6c–d**) reinforce these observations. $AoI_1$ displays a monotonic profile with a single peak and smooth decay, whereas $AoI_2$ exhibits non-monotonic behavior with multiple peaks due to AAo-DA interaction. Overall, the flow in the normal fetal aortic arch is predominantly planar, with weak but persistent 3D effects. The DA-induced secondary jet locally enhances near-wall velocity gradients, which are directly relevant to WSS evaluation.

Peak velocities are concentrated near the lower wall. Along $AoI_1$, a single dominant peak is observed, corresponding to the AAo inflow. At $AoI_2$, a secondary peak emerges due to the interaction between the DA jet and the primary stream, resulting in a non-monotonic velocity distribution with multiple local extrema.

### C.3.2 Flow Field in the Coarcted Fetal Aorta and its 3D Effect

Compared with the normal aorta, the coarcted configuration exhibits pronounced differences in both velocity contours [**Fig. 7(a–b)**] and profiles [**Fig. 7(c–d)**]. Immediately downstream of the cross-section line $AoI_1$, flow separation develops along the lower wall. The circled region in **Fig. 7(b)** highlights this separated zone, which arises from an adverse pressure gradient downstream of the constriction. Reattachment occurs further downstream, near $AoI_2$, and is promoted by interaction with the DA inflow, which injects additional momentum into the near-wall region. Consequently, this region experiences a transition from an adverse to a favorable pressure gradient.

Velocity profiles at $AoI_1$ and $AoI_2$ [**Fig. 7(c–d)**] reveal substantial deviations from the normal configuration. At $AoI_1$, peak velocity increases markedly and shifts toward the wall adjacent to the constriction, producing a steep wall-normal gradient. At $AoI_2$, the profile contains multiple local extrema, reflecting the combined effects of jet breakdown, flow reattachment, and interaction with the DA jet. The secondary peak is displaced relative to the normal case, indicating a modified flow topology within the arch.

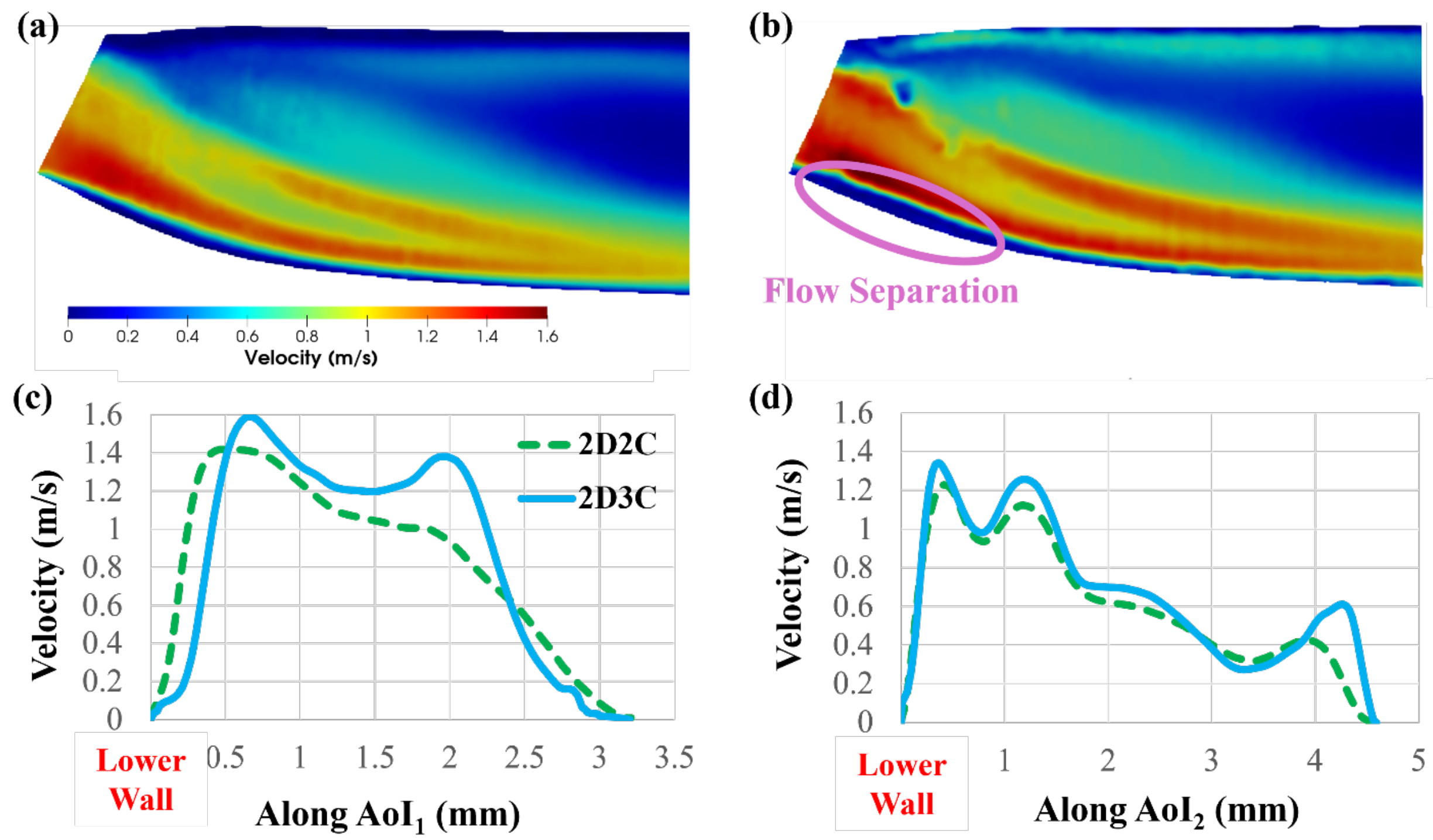

Figure 7 Velocity field in the coarcted fetal aortic arch obtained from (a) 2D2C and (b) 2D3C PIV, showing jet acceleration, flow separation, and reattachment downstream of the coarctation. Velocity profiles at cross-section lines, (c) $AoI_1$ and (d) $AoI_2$, highlight peak amplification, wallward shift of high-speed flow, and increased three-dimensionality compared with the normal configuration. AoI: aortic isthmus.

The 2D3C measurements further reveal appreciable out-of-plane velocity components, indicating strong three-dimensionality in the coarcted flow field. In contrast, 2D2C captures only the in-plane projection and under-represents the true velocity magnitude and gradients, especially in the vicinity of the constriction. These findings are consistent with continuity: the reduced cross-sectional area at the coarctation throat increases local velocity, sharpens near-wall gradients, and shifts the location of maximum velocity toward the vessel wall, thereby enhancing flow asymmetry and spatial heterogeneity.

Overall, whereas the normal configuration remains predominantly planar with modest 3D effects, the coarcted geometry generates pronounced 3D flow and significantly alters the near-wall velocity field. These changes directly affect the magnitude and distribution of WSS, highlighting the limitations of relying solely on 2D2C measurements in stenotic geometries.

#### C.3.3 Wall Shear Stress and its 3D Effect

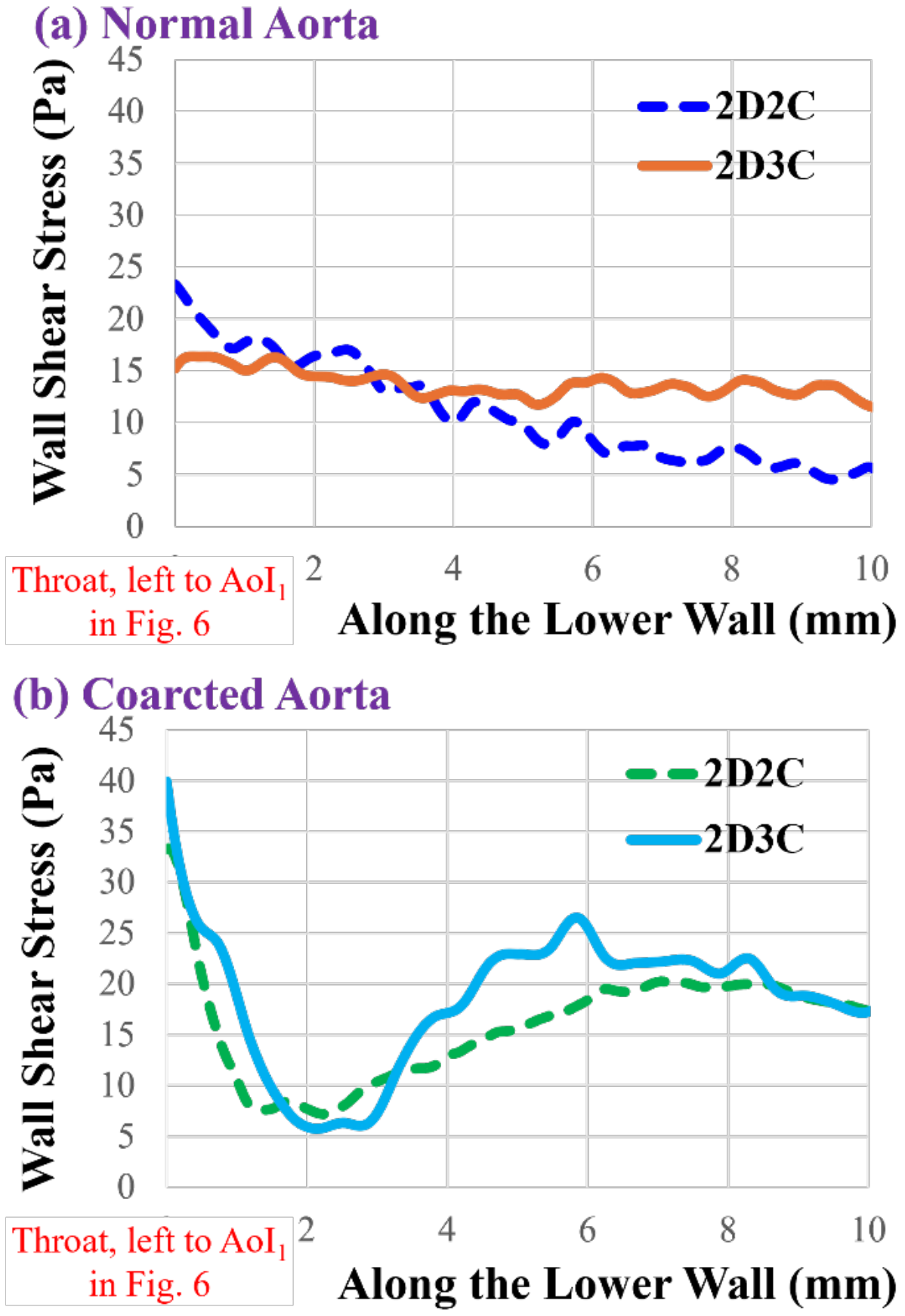

Figure 8 Wall shear stress distributions along the lower wall comparing 2D2C and 2D3C measurements in the (a) normal and (b) coarcted aorta.

WSS distributions along the lower wall, derived from both 2D2C and 2D3C PIV measurements (**Fig. 8**), reveal clear three-dimensional effects in both normal and coarcted fetal aortas. In all cases, 2D2C systematically underestimates WSS relative to 2D3C, with the discrepancy being more pronounced in the coarcted configuration. At the coarctation throat, WSS is markedly amplified compared with the normal aorta, reflecting strong contraction-induced acceleration and steepened near-wall velocity gradients.

Immediately downstream of the throat (≈1–2 mm), WSS drops sharply in both measurement configurations, consistent with flow separation driven by an adverse pressure gradient and a corresponding reduction in wall-normal velocity gradients. Further downstream (≈3–7 mm), WSS increases again as the flow reattaches under a favorable pressure gradient, and beyond ≈8 mm the 2D2C and 2D3C profiles gradually converge, indicating partial redevelopment of the velocity field.

When integrated along the lower wall, WSS obtained from 2D2C is approximately 17% lower than 2D3C in the normal aorta and about 29% lower in the coarcted case. Although the overall spatial patterns are qualitatively similar, the magnitude underestimation is greatest in regions of elevated shear near the constriction. These findings demonstrate that neglecting out-of-plane velocity components leads to substantial bias in WSS estimation, particularly in strongly 3D stenosis-driven flows.

### C.4 Comparison Between Normal and Coarcted Fetal Aortas

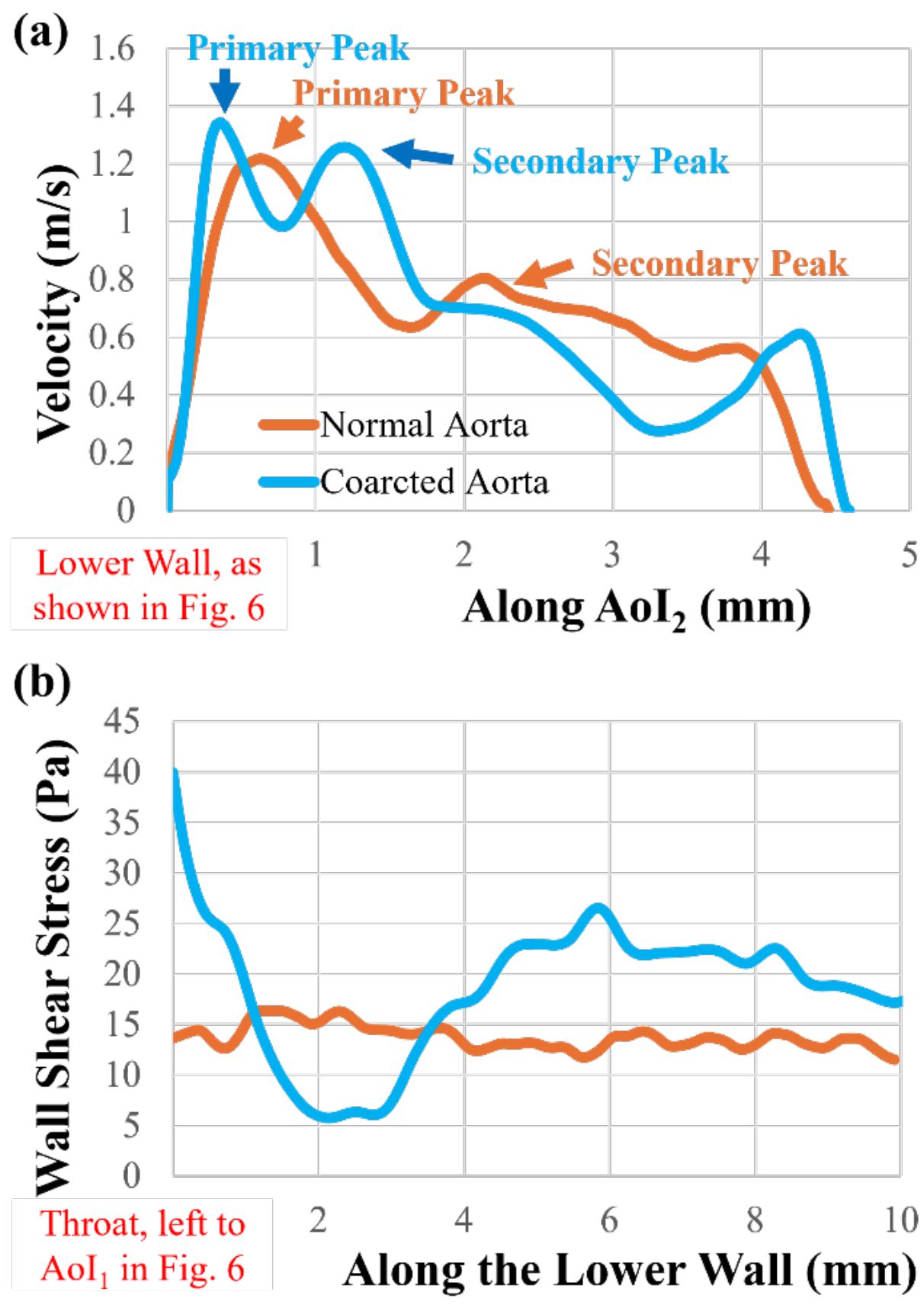

Figure 9 Summary comparison of hemodynamics at $AoI_2$ between normal and coarcted fetal aortas based on 2D3C PIV data. Velocity profiles in (a) reproduce and juxtapose $AoI_2$ profiles extracted from Figs. 6 and 7, highlighting the two-peak structure in both cases and the wall-directed amplification of the secondary jet in the coarcted geometry. Wall shear stress distributions in (b) synthesize WSS results from Figs. 7 and 8, showing pronounced WSS amplification at the coarctation throat, a downstream low-shear separation region, and gradual recovery along the lower wall. AoI: aortic isthmus.

The 2D3C velocity profiles at $AoI_2$ for the normal and coarcted geometries both exhibit a characteristic two-peak structure (**Fig. 9a**). In the normal aorta, the secondary peak remains within the core flow and is weaker than the primary peak, indicating a more distributed momentum field with limited direct wall interaction. In the coarcted aorta, however, the secondary jet is deflected toward the lower wall and increases in magnitude, demonstrating that constriction-induced acceleration redistributes momentum asymmetrically and generates a wall-directed jet.

This jet deflection is consistent with previous observations in stenosed aortic flows, where eccentric high-velocity jets impinge on the vessel wall, creating localized regions of elevated WSS.[52] Such eccentric or displaced flow structures have been linked to altered hemodynamic loading and aortic remodeling[53], and abnormal WSS is known to modulate endothelial function and contribute to the development of vascular disease[54]. Consequently, the wall-directed jet observed in the coarcted configuration provides a mechanistic explanation for the formation of spatially concentrated shear "hot spots" that may predispose the aortic wall to adverse adaptive or maladaptive remodeling.

The corresponding WSS distributions along the lower wall (**Fig. 9b**) further clarify these effects and delineate three distinct shear regions. First, a localized high-shear zone is present at the coarctation throat, where geometric narrowing and jet acceleration dramatically steepen the near-wall velocity gradient. Second, a downstream low-shear region emerges, associated with flow separation and displacement of the jet core away from the wall. Third, a recovery region is observed further downstream, where reattachment and redevelopment of the boundary layer restore higher wall-normal gradients. The persistence of elevated and spatially heterogeneous WSS beyond the throat indicates that the hemodynamic influence of the coarctation extends over a substantial portion of the arch, with important implications for site-specific patterns of vascular remodeling and potential long-term sequelae.

### C.5 Limitations and Future Work

The present study establishes a controlled experimental framework for quantifying velocity fields and WSS in a patient-specific fetal aortic arch; however, several limitations should be acknowledged. The aortic phantom was modeled as rigid, whereas the fetal aorta is inherently compliant, and vessel deformation can alter flow structure and near-wall velocity gradients, thereby affecting WSS magnitude and distribution. Also, experiments were performed under steady-flow conditions, while fetal circulation is intrinsically pulsatile. Consequently, temporal variations in flow and shear were not captured. Future work will incorporate physiologically representative pulsatile waveforms to improve physiological fidelity and to further quantify measurement-induced bias in WSS estimation.

Moreover, the current analysis is also limited to a single subject-specific geometry, which constrains generalization across the broader spectrum of fetal aortic anatomies. Future studies will therefore extend the framework to a larger cohort of subject-specific fetal aortic models to assess inter-subject variability in key hemodynamic metrics and to evaluate the generality of WSS calculation across anatomically diverse configurations.

Despite these limitations, this first-of-its-kind study demonstrates the feasibility of constructing a human-based fetal aortic in vitro platform and resolving detailed hemodynamics using microscale stereoscopic PIV, providing a foundation for more physiologically comprehensive investigations.

## D. CONCLUSIONS

This study introduced a first-of-its-kind, human-based in vitro platform for quantifying fetal aortic hemodynamics under normal and coarctation conditions. An image-based fetal aortic phantom integrated into a dual-inflow flow loop reproduced target clinical flow distributions and, combined with high-quality optics and a characterized working fluid, enabled high-resolution planar and stereoscopic PIV measurements.

In the normal configuration, flow was predominantly planar with modest 3D effects, whereas coarctation produced marked jet acceleration, separation, reattachment, and strongly heterogeneous WSS extending beyond the geometric narrowing. Two-component PIV systematically underestimated WSS relative to three-component measurements (by ~17% in the normal aorta and ~29% in the coarcted case), especially in high-shear, strongly 3D regions.

These findings highlight the necessity of three-component measurements for accurate WSS estimation in complex fetal aortic geometries and suggest that detailed hemodynamic metrics, such as WSS, may serve as potential biomarkers to improve fetal CoA diagnosis beyond anatomy-based assessment alone.

## E. ACKNOWLEDGMENTS

This work is supported by the Saving Tiny Hearts Society, American Heart Association [AHA-23SCEFIA1157994], and National Institute of Biomedical Imaging and Bioengineering [NIBIB-R21EB034833],

## F. DECLARATION OF COMPETING INTEREST

The authors declare that they have no known competing financial interests or personal relationships that could have appeared to influence the work reported in this paper.

## G. AUTHOR CONTRIBUTIONS

- Kamlesh Kumar: Data curation (lead); Formal analysis (lead); Investigation (lead); Validation (lead); Visualization (lead); Writing – original draft (lead); Writing – review & editing (equal).
- Biao Si: Data curation (equal); Formal analysis (equal); Investigation (equal); Methodology (supporting); Validation (equal); Visualization (equal); Writing – review & editing (equal).
- Pritom Saha: Data curation (supporting); Formal analysis (supporting); Investigation (supporting); Methodology (supporting); Resources (supporting); Writing – review & editing (supporting).
- Shuping Ge: Conceptualization (equal); Funding acquisition (supporting); Investigation (supporting); Methodology (supporting); Project administration (equal); Resources (equal); Supervision (equal); Writing – review & editing (supporting).
- Zhenglun Alan Wei: Conceptualization (lead); Formal analysis (supporting); Funding acquisition (lead); Methodology (lead); Project administration (lead); Resources (lead); Supervision (lead); Validation (equal); Writing – original draft (equal); Writing – review & editing (equal).

## H. DATA AVAILABILITY

Data supporting the findings of this study can be obtained from the corresponding author upon reasonable request.